\documentclass[twoside,twocolumn,9pt]{article}
\usepackage{extsizes}
\usepackage[super,sort&compress,comma]{natbib} 
\usepackage[version=3]{mhchem}
\usepackage[left=1.5cm, right=1.5cm, top=1.785cm, bottom=2.0cm]{geometry}
\usepackage{hyperref}
\hypersetup{colorlinks=true,linkcolor=black,citecolor=black}
\usepackage{balance}
\usepackage{mathptmx}
\usepackage{sectsty}
\usepackage{graphicx} 
\usepackage{lastpage}
\usepackage[format=plain,justification=justified,singlelinecheck=false,font={stretch=1.125,small,sf},labelfont=bf,labelsep=space]{caption}
\usepackage{float}
\usepackage{fancyhdr}
\usepackage{fnpos}
\usepackage[english]{babel}
\addto{\captionsenglish}{%
  \renewcommand{\refname}{Notes and references}
}
\usepackage{array}
\usepackage{droidsans}
\usepackage{charter}
\usepackage[T1]{fontenc}
\usepackage[usenames,dvipsnames]{xcolor}
\usepackage{setspace}
\usepackage[compact]{titlesec}
\usepackage{hyperref}

\usepackage{epstopdf}

\definecolor{cream}{RGB}{222,217,201}

\begin{document}

\pagestyle{fancy}
\thispagestyle{plain}
\fancypagestyle{plain}{
\renewcommand{\headrulewidth}{0pt}
}

\makeFNbottom
\makeatletter
\renewcommand\LARGE{\@setfontsize\LARGE{15pt}{17}}
\renewcommand\Large{\@setfontsize\Large{12pt}{14}}
\renewcommand\large{\@setfontsize\large{10pt}{12}}
\renewcommand\footnotesize{\@setfontsize\footnotesize{7pt}{10}}
\renewcommand\scriptsize{\@setfontsize\scriptsize{7pt}{7}}
\makeatother

\renewcommand{\thefootnote}{\fnsymbol{footnote}}
\renewcommand\footnoterule{\vspace*{1pt}%
\color{cream}\hrule width 3.5in height 0.4pt \color{black} \vspace*{5pt}} 
\setcounter{secnumdepth}{5}

\makeatletter 
\renewcommand\@biblabel[1]{#1}            
\renewcommand\@makefntext[1]%
{\noindent\makebox[0pt][r]{\@thefnmark\,}#1}
\makeatother 
\renewcommand{\figurename}{\small{Fig.}~}
\sectionfont{\sffamily\Large}
\subsectionfont{\normalsize}
\subsubsectionfont{\bf}
\setstretch{1.125} 
\setlength{\skip\footins}{0.8cm}
\setlength{\footnotesep}{0.25cm}
\setlength{\jot}{10pt}
\titlespacing*{\section}{0pt}{4pt}{4pt}
\titlespacing*{\subsection}{0pt}{15pt}{1pt}

\fancyfoot{}
\fancyfoot[LO,RE]{\vspace{-7.1pt}\includegraphics[height=9pt]{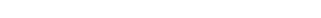}}
\fancyfoot[CO]{\vspace{-7.1pt}\hspace{13.2cm}\includegraphics{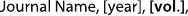}}
\fancyfoot[CE]{\vspace{-7.2pt}\hspace{-14.2cm}\includegraphics{head_foot/RF}}
\fancyfoot[RO]{\footnotesize{\sffamily{1--\pageref{LastPage} ~\textbar  \hspace{2pt}\thepage}}}
\fancyfoot[LE]{\footnotesize{\sffamily{\thepage~\textbar\hspace{3.45cm} 1--\pageref{LastPage}}}}
\fancyhead{}
\renewcommand{\headrulewidth}{0pt} 
\renewcommand{\footrulewidth}{0pt}
\setlength{\arrayrulewidth}{1pt}
\setlength{\columnsep}{6.5mm}
\setlength\bibsep{1pt}

\makeatletter 
\newlength{\figrulesep} 
\setlength{\figrulesep}{0.5\textfloatsep} 

\newcommand{\topfigrule}{\vspace*{-1pt}%
\noindent{\color{cream}\rule[-\figrulesep]{\columnwidth}{1.5pt}} }

\newcommand{\botfigrule}{\vspace*{-2pt}%
\noindent{\color{cream}\rule[\figrulesep]{\columnwidth}{1.5pt}} }

\newcommand{\dblfigrule}{\vspace*{-1pt}%
\noindent{\color{cream}\rule[-\figrulesep]{\textwidth}{1.5pt}} }

\makeatother

\twocolumn[
  \begin{@twocolumnfalse}
{\includegraphics[height=30pt]{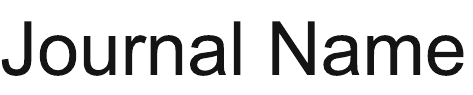}\hfill\raisebox{0pt}[0pt][0pt]{\includegraphics[height=55pt]{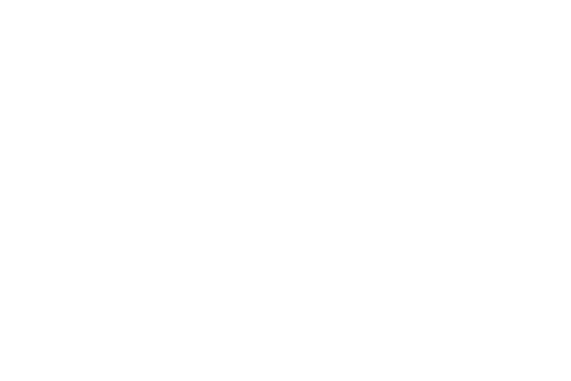}}\\[1ex]
\includegraphics[width=18.5cm]{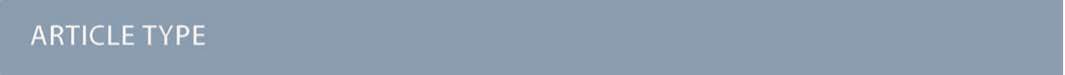}}\par
\vspace{1em}
\sffamily
\begin{tabular}{m{4.5cm} p{13.5cm} }

\includegraphics{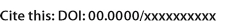} & \noindent\LARGE{\textbf{High resistance of superconducting TiN thin films against environmental attacks$^\dag$}} \\
 & \vspace{0.3cm} \\

 & \noindent\large{Zhangyuan Guo,\textit{$^{a,b}$}$^{\ddag}$ Min Ge,\textit{$^{c}$}$^{\ddag}$ You-Qi Zhou,\textit{$^{a,d}$}$^{\ddag}$ Jiachang Bi,$^{\ast}$\textit{$^{b,e}$} Qinghua Zhang,\textit{$^{f}$} Jiahui Zhang,\textit{$^{b}$} Jin-Tao Ye,\textit{$^{d}$} Rongjing Zhai,\textit{$^{b}$} Fangfang Ge,\textit{$^{b}$} Yuan Huang,\textit{$^{g}$} Ruyi Zhang,\textit{$^{b}$} Xiong Yao,\textit{$^{b}$} Liang-Feng Huang,$^{\ast}$\textit{$^{d}$}and Yanwei Cao$^{\ast}$\textit{$^{b,e}$}
 } \\

\includegraphics{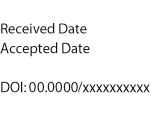} & \\

\end{tabular}

 \end{@twocolumnfalse} \vspace{0.6cm}

  ]

\renewcommand*\rmdefault{bch}\normalfont\upshape
\rmfamily
\section*{}
\vspace{-1cm}


\footnotetext{\textit{$^{a}$~ School of Materials Science and Chemical Engineering, Ningbo University, Ningbo, Zhejiang 315211, China. }}
\footnotetext{\textit{$^{b}$~Ningbo Institute of Materials Technology and Engineering, Chinese Academy of Sciences, Ningbo 315201, China. E-mail: bijiachang@nimte.ac.cn, ywcao@nimte.ac.cn }}
\footnotetext{\textit{$^{c}$~The Instruments Center for Physical Science, University of Science and Technology of China, Hefei 230026, China}}
\footnotetext{\textit{$^{d}$~Research Center for Advanced Interdisciplinary Sciences, Ningbo Institute of Materials Technology and Engineering, Chinese Academy of Sciences, Ningbo 315201, China. E-mail: huangliangfeng@nimte.ac.cn}}
\footnotetext{\textit{$^{e}$Center of Materials Science and Optoelectronics Engineering, University of Chinese Academy of Sciences, Beijing 100049, China }}
\footnotetext{\textit{$^{f}$Beijing National Laboratory for Condensed Matter Physics, Institute of Physics, Chinese Academy of Sciences, Beijing 100190, China}}
\footnotetext{\textit{$^{g}$Advanced Research Institute of Multidisciplinary Science, Beijing Institute of Technology, Beijing 100081, China}}
\footnotetext{\dag~Electronic Supplementary Information (ESI) available.}
\footnotetext{\ddag~These authors contributed equally to this work.}



\sffamily{\textbf{Superconductors, an essential class of functional materials, hold a vital position in both fundamental science and practical applications. However, most superconductors, including MgB$_2$, Bi$_2$Sr$_2$CaCu$_2$O$_{8+\delta}$, and FeSe, are highly sensitive to environmental attacks (such as water and moist air), hindering their wide applications. More importantly, the surface physical and chemical processes of most superconductors in various environments remain poorly understood. Here, we comprehensively investigate the high resistance of superconducting titanium nitride (TiN) epitaxial films against acid and alkali attacks. Unexpectedly, despite immersion in acid and alkaline solutions for over 7 days, the crystal structure and superconducting properties of TiN films remain stable, as demonstrated by high-resolution X-ray diffraction, electrical transport, atomic force microscopy, and scanning electron microscope.  Furthermore, combining scanning transmission electron microscopy analysis with density functional theory calculations revealed the corrosion mechanisms: acid corrosions lead to the creation of numerous defects due to the substitution of Cl ions for N anions, whereas alkaline environments significantly reduce the film thickness through the stabilization of OH$^\ast$ adsorbates. Our results uncover the unexpected stability and durability of superconducting materials against environmental attacks, highlighting their potential for enhanced reliability and longevity in diverse applications.}}\\


\rmfamily 


\section{Introduction}
\begin{figure*}[t]
	\centering
	\includegraphics[height=14cm]{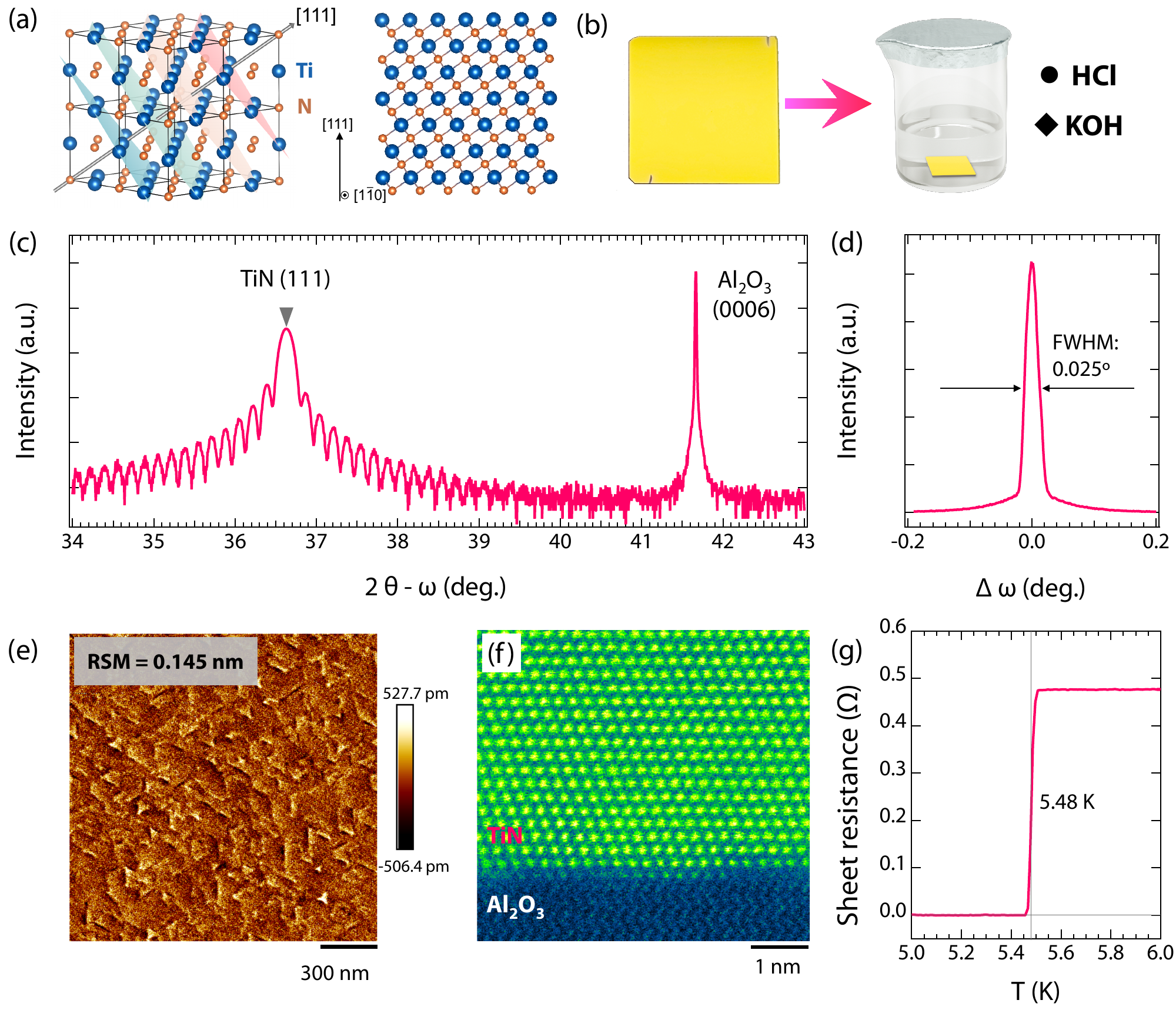}
	\caption{(a) The schematic of the TiN crystal structure with a rock-salt
		symmetry (left panel) and the side view of (111)-oriented TiN crystal structure (right panel). (b) Photographs of the pristine TiN thin films (left panel) and a schematic of the TiN films immersed in HCl and KOH solutions (right panel). (c) 2$\theta$-$\omega$ scans around the (111) diffraction of pristine TiN films. (d) Rocking curve and (e) surface morphology of pristine TiN films. (f) HAADF-STEM image of pristine TiN films at the atomic scale. (g) Temperature-dependent sheet resistance of pristine TiN films near the superconducting transition temperature (T$\rm_c$ $\sim$ 5.48 K).} 
	\label{figure1}
\end{figure*}

Superconductors, a significant class of functional materials with zero electrical resistance and the Meissner effect, play an essential role in both fundamental science and practical applications. They are employed in fields such as quantum computing, medical imaging, communication, radar systems, particle accelerators, and the development of highly efficient power grids and maglev trains, revolutionizing technology and energy efficiency. \cite{Science-2021-Leon, MRS-2013-Oliver, AnnRev-2012-Zmuidzinas, IScience-2021-Yao,RDMS-2018-Gupta} In various applications, the stability of superconductivity ensures reliable and efficient performance. Generally, superconductors are metals or compounds. On the one hand, most metal (alloy) superconductors, such as Nb, Ta, and NbTi, are chemically stable, yet the study of their superconductivity stability is lacking. \cite{Corrosion-1958-Hampel,Niobium-2001-Graham,ASM-2005-Sutherlin,MaterSci-2011-Bai, MRS-2011-Delacour} On the other hand, most compound superconductors, such as oxides, sulfides, and borides, particularly high-temperature superconductors, are unstable against environmental attacks. \cite{APL-1987-Yan,SST-1988-Liu,APL-1993-Zhou,MRS-1993-Barkatt,JMPT-2007-Argyropoulou,Nature-2019-Yu, ACSAMI-2022-Huang, SciChina-2022-Ding,JAC-2024-Zhao} For instance, the superconducting transition temperature of both the copper oxide high-temperature superconductors Bi$_2$Sr$_2$CaCu$_2$O$_{8+\delta}$ and the nickelate superconductor Nd$_{0.8}$Sr$_{0.2}$NiO$_2$ decreases significantly when exposed to water. \cite{ACSAMI-2022-Huang, SciChina-2022-Ding} The intermetallic compound superconductors Nb$_3$Sn dissolve in NaCl solutions, \cite{MIT-2015-Priyotomo} and MgB$_2$ decomposes when exposed to water. \cite{JMPT-2007-Argyropoulou,SST-2001-Zhai} In the case of chalcogenide superconductors, such as NbSe$_2$, they are generally layered and unstable in air. \cite{NanoLetter-2015-Cao} The high-temperature superconductor FeSe can oxidize rapidly when exposed to air, \cite{PRB-2009-Pomjakushina} and water has detrimental effects on the iron-based superconductor Ba$_{0.6}$K$_{0.4}$Fe$_2$As$_2$. \cite{IEEE-20214-Tu} Meanwhile, considerable effort has been devoted to developing environmentally stable chalcogenide superconductors. \cite{NM-2019-Lin,SciAdv-2023-Song} Hence, it is crucial to investigate the surface physical and chemical processes of superconductors in various environments to explore the stability and durability of superconducting materials.

\begin{figure*}[t]
	\centering
	\includegraphics[height=10cm]{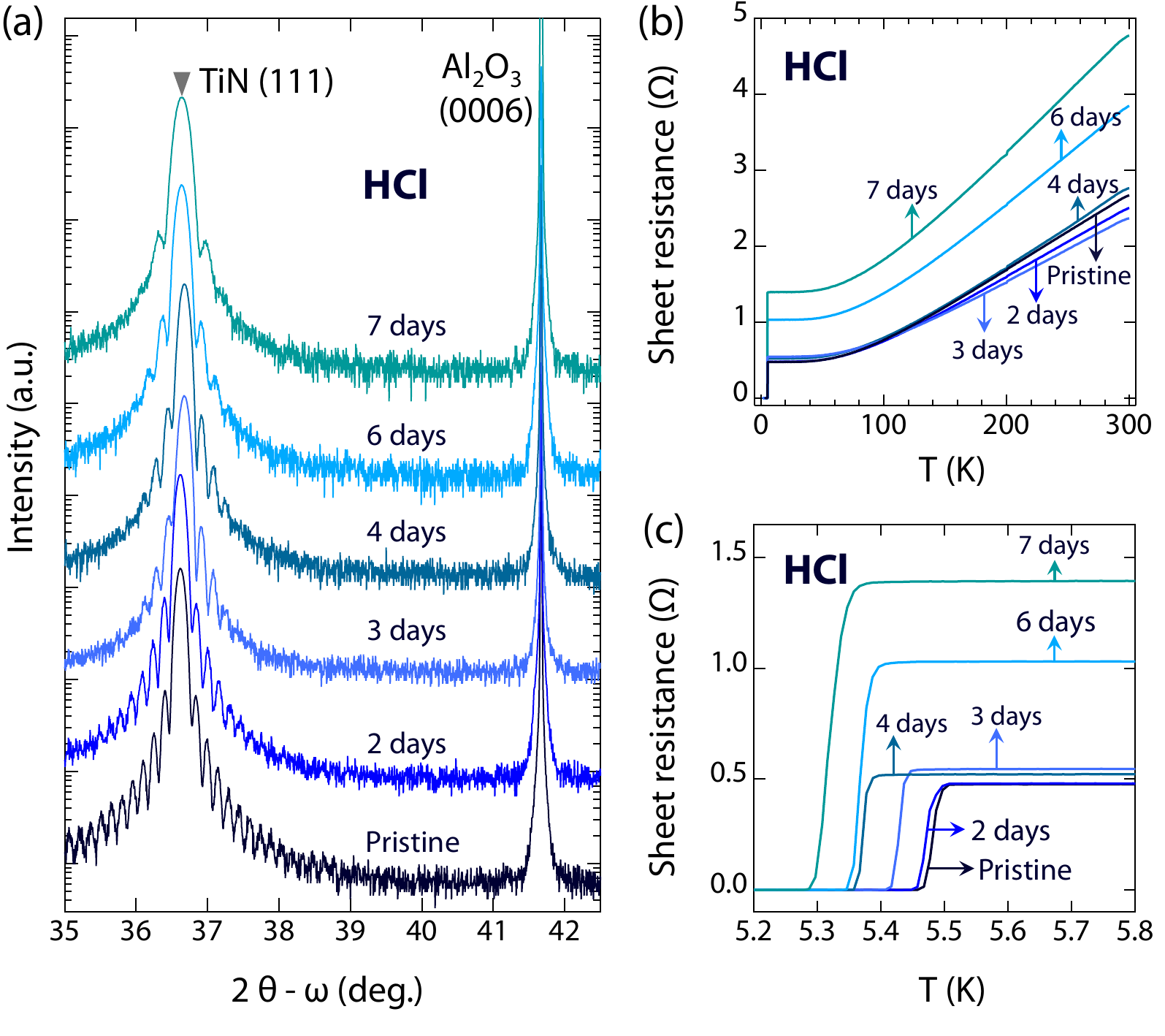}
	\caption{(a) 2$\theta$-$\omega$ scans around the (111) diffraction peaks, and (b) temperature-dependent sheet resistances of pristine TiN films and TiN films immersed in HCl for 2, 3, 4, 6, and 7 days, respectively. (c) The enlarged view of (b) around T$\rm_c$. }
	\label{figure2}
\end{figure*}

Among various superconducting materials, transition metal nitrides are well-known for their physical and chemical stability. Titanium nitride (TiN), a representative of them, has been extensively used as a hard and anti-corrosion coating in harsh environments due to its remarkable chemical, mechanical, and thermal stability, as well as its wear-resistant properties. \cite{ThinSolidFoilms-1985-Sundgren,SCI-2019-Caha,MaterCorr-2019-Rodrigues,SNAS-2020-Sarkar, DMJ-2018-Sugisawa, JPD-2009-Mahieu, JMagAlloy-2014-Tacikowski, ACSMaterLett-2024-Silva, ASS-2022-Ghailane, CeramInt-2022-Lin, SurfCoatInt-2019-Caha} Moreover, its excellent electrical conductivity and CMOS compatibility have led to its widespread use in semiconductor devices. \cite{SciRep-2014-Chen, IEEE-2004-Chau,ACSAMI-2022-Ho, ACSNano-2023-Li, Nanoscale-2023-Sun} Very recently, TiN was utilized as transparent electrodes with resistance to acid and alkali, overcoming the corrosion challenges of traditional transparent electrode materials (e.g., indium tin oxide). \cite{ACSNano-2023-Li} Interestingly, TiN has emerged as a promising material for superconducting quantum computers due to its ultrahigh quality factors, long qubit coherence time, and high kinetic inductance. \cite{Science-2021-Leon,JAP-2012-Krockenberger,APL-2010-Vissers,APL-2013-Chang,APL-2018-Shearrow,APL-2020-Melville,PRM-2022-Gao,AM-2024-Gao,PRA-2023-Deng} Its heteroepitaxy with topological insulators (such as Bi$_2$Se$_3$) further increases the potential of superconducting TiN for topological quantum computing. \cite{ACSAMI-2024-Xie} However, the resistance of superconducting single-crystalline TiN thin films against environmental attacks remains poorly understood at present. 

In this work, a series of single-crystalline TiN (111) films, synthesized on $\alpha$-Al$_2$O$_3$ (0001) substrates by a homemade magnetron sputtering epitaxy, \cite{PRM-2021-Bi,NanoLetter-2024-Bi} were immersed in strong acid and alkaline solutions to investigate the surface physical and chemical processes. The high resistance of superconducting TiN thin films against environmental attacks was characterized through high-resolution X-ray diffraction (XRD) and electrical transport measurements. Atomic force microscopy (AFM), scanning electron microscope (SEM), and scanning transmission electron microscopy (STEM) revealed remarkably different corrosion processes in acidic and alkaline solutions. Furthermore, the corrosion mechanisms were uncovered by combining STEM and density functional theory (DFT) calculations. Our results uncover the micromechanism of superconducting TiN thin films against environmental attacks.

\begin{figure*}[t]
	\centering
	\includegraphics[height=10cm]{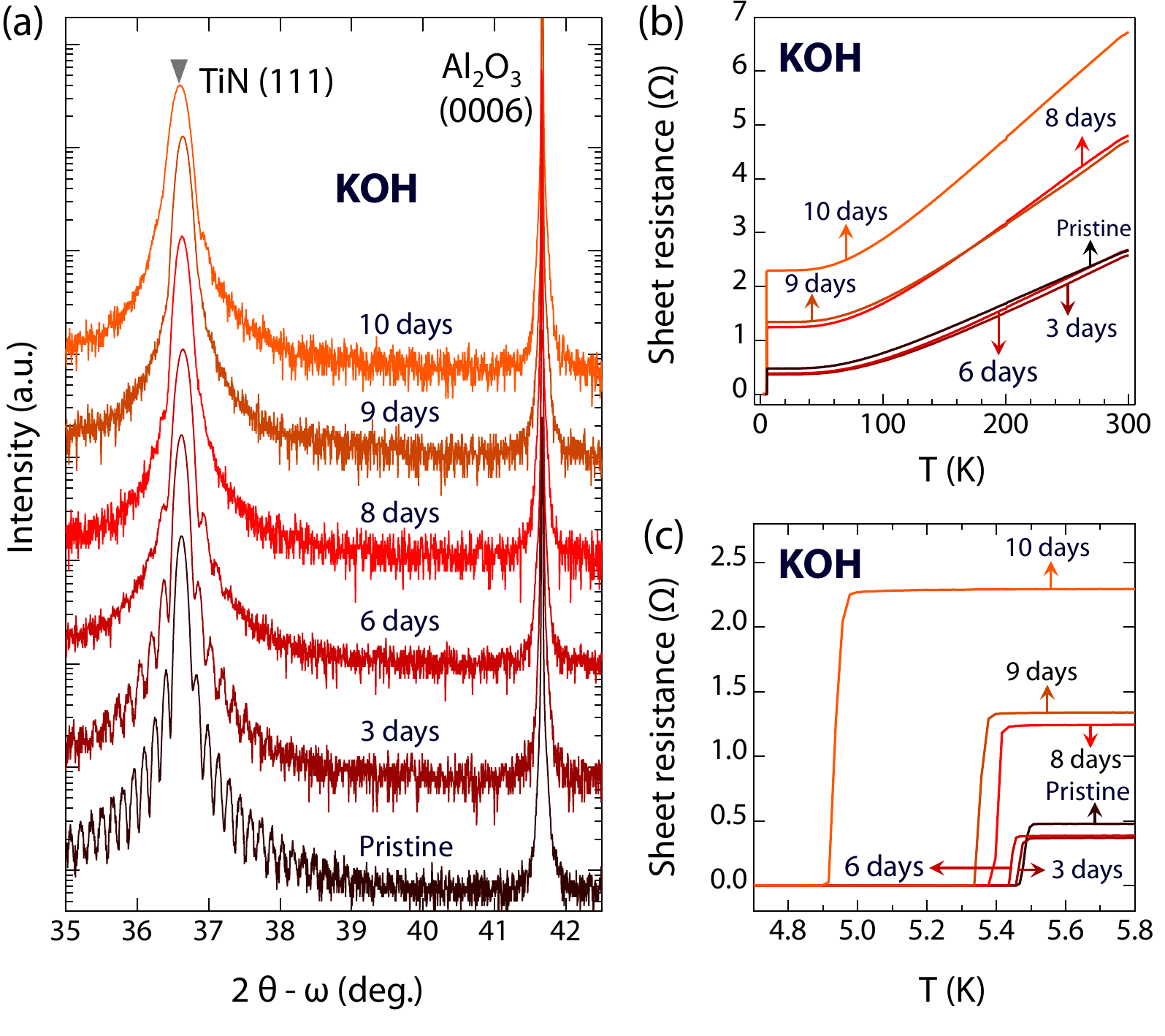}
	\caption{(a) 2$\theta$-$\omega$ scans around the (111) diffraction peaks, and (b) temperature-dependent sheet resistances of pristine TiN films and TiN films immersed in KOH for 3, 6, 8, 9, and 10 days, respectively. (c) The enlarged view of (b) around T$\rm_c$.}
	\label{figure3}
\end{figure*}

\section{Results and Discussion}

Figure 1 shows the crystal structures of TiN (111) films. As seen in Fig.\ref{figure1}(a), bulk TiN has a cubic rocksalt structure with a space group of $ Fm\bar{3}m$. Interestingly, the specific (111)-oriented TiN, consisting of alternate Ti and N atomic layers (see Fig.\ref{figure1}(a)), enables the high resistance to oxidation due to the fact that the exposure of N atomic layers tends to inhibit oxygen adsorption. \cite{ACSPhotonics-2021-Zhang} Importantly, single-crystalline TiN (111) films show more superior corrosion resistance than both polycrystalline TiN films and single-crystalline TiN (001) films, as shown in Fig. S1 and Fig. S2 (ESI\dag). Therefore, to investigate the resistance of superconducting TiN thin films against environmental attacks, high-quality TiN (111) films with bright golden color (see Fig.\ref{figure1}(b)) were synthesized on $\alpha$-Al$_2$O$_3$ (0001) substrates. As seen in Fig.\ref{figure1}(c), the 2$ \theta-\omega $ scans show a dominative (111) diffraction peak without any detectable secondary phase, indicating the epitaxial growth of TiN thin films. Furthermore, the distinct thickness fringes in the 2$ \theta-\omega $ scan (2$\theta$ between 34 and 39$^\circ$, Fig.\ref{figure1}(c)), the narrow rocking curve ($\sim$ 0.025$^\circ$, Fig.\ref{figure1}(d)), and the small surface roughness ($\sim$ 0.145 nm, Fig.\ref{figure1}(e)) all demonstrate the smooth surface morphology and high quality of TiN thin films. Figure \ref{figure1}(f) shows the HAADF-STEM image of TiN thin films, displaying the crystal structures at the atomic scale. As seen, the TiN thin films are highly single-crystalline with an atomically sharp interface between the film and the substrate, further verifying the high quality of the films. Moreover, a sharp transition from the normal to superconducting states presents at 5.48 K (see Fig.\ref{figure1}(g)). This transition temperature is comparable to that of TiN films grown by plasma enhanced molecular beam epitaxy (5.25 -5.4 K) \cite{JAP-2020-Richardson,JAP-2012-Krockenberger} and is close to the transition temperature of bulk TiN (T$\rm_c$ = 5.6 K). \cite{RMP-1963-Matthias,PR-1954-Hardy} Both the high T$\rm_c$ and sharp superconducting transition ($\sim$ 0.05 K, from 5.51 to 5.46 K) further confirm the high quality of TiN films. 

\begin{figure}[h]
	\centering
	\includegraphics[height=7.5cm]{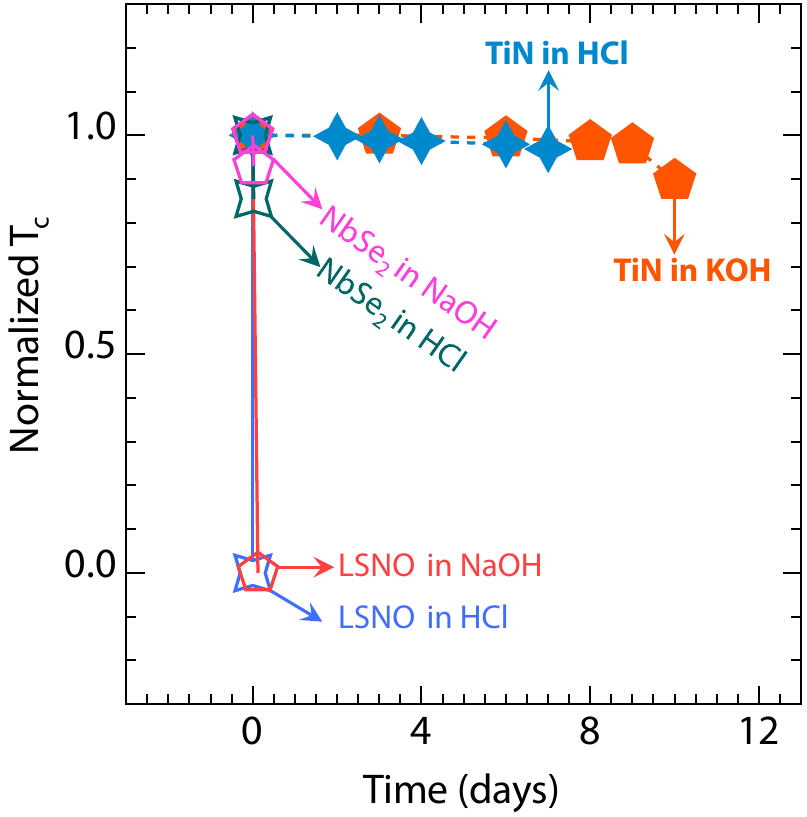}
	\caption{The T$\rm_c$ summary of TiN films, La$_{0.8}$Sr$_{0.2}$NiO$_2$ films\cite{JAC-2024-Zhao}, and NbSe$_2$ films\cite{NM-2019-Lin} in acidic and alkaline environments.}
	\label{figure4}
\end{figure}

To investigate the resistance of superconducting TiN films in acid and alkaline environments, a series of high-quality TiN (111) films were immersed in 1 mol/L HCl (pH $\sim$ 1) and 1 mol/L KOH (pH $\sim$ 14), respectively. Figure \ref{figure2}(a) shows the evolution of the 2$\theta$-$\omega$ scans around the (111) diffraction of TiN films immersed in HCl for different durations (0, 2, 3, 4, 6, and 7 days). As seen, the positions of the (111) diffraction peaks of the TiN films have no significant changes, even after immersion for 7 days (2$\theta$ is 36.638$^\circ$, compared to 36.618$^\circ$ for pristine TiN films), indicating the high stability of the crystal structures of TiN films in the HCl environment. Despite of the single crystalline feature, the thickness fringes become weaker with extended immersion time, indicating the increased surface roughness of the films. Besides the crystal structures, we also explore the evolution of the electronic structures. Figure \ref{figure2}(b) shows temperature-dependent sheet resistances of TiN films immersed in HCl. As seen, the sheet resistance of TiN films is almost unchanged, between 2.36 and 2.76 $\Omega$, when immersed in HCl for up to 4 days. However, the sheet resistance gradually increases to 3.85 $\Omega$ after 6 days and to 4.77 $\Omega$ after 7 days of immersion. The slightly increased sheet resistance may be due to the increasing defects caused by HCl corrosion. It is noteworthy that the T$\rm_c$ of TiN films immersed in HCl remains almost unchanged or slightly decreases compared to pristine TiN films, as shown in Fig.\ref{figure2}(c) and Fig.\ref{figure4}. The T$\rm_c$ of TiN films immersed in HCl for 7 days (5.32 K) shows only a slight decrease compared to pristine TiN films (5.48 K). The reduction in T$\rm_c$ can be negligible, and the critical current density of TiN films immersed in HCl for 6 days is close to that of pristine TiN films (see Fig.S3, ESI\dag), indicating the high resistance of superconducting TiN films against acid corrosion.

In addition to acidic environments, next we study the resistance of superconducting TiN films in alkaline conditions. Figure \ref{figure3} shows the evolution of the crystal structure and superconductivity of films immersed in KOH. Similar to the results for TiN films immersed in HCl, the positions of the (111) diffraction peaks of the TiN films are also almost unchanged, while the thickness fringes disappear with extended immersion time, as shown in Fig. \ref{figure3}(a). These observations indicate that the single crystalline feature of the TiN films survives during KOH corrosion, whereas the flat surface was heavily destroyed. For the electronic structures, the sheet resistance of TiN films immersed in KOH remains nearly constant at approximately 2.6 $\Omega$ within 6 days. The slightly decreased sheet resistances of TiN films immersed in KOH for 3 and 6 days can result from the individual differences among pristine TiN films with the same growth conditions (see Fig.S4, ESI\dag). However, the sheet resistance begins to increase with further immersion, reaching 4.8 $\Omega$ after 8 and 9 days, and up to 6.7 $\Omega$ after 10 days. Regarding the special interest in superconductivity, the superconductivity of TiN films is robust even immersed in KOH for up to 9 days (T$\rm_c$ is 5.36 K), as shown in Fig.\ref{figure3}(c) and Fig.\ref{figure4}.  After 10 days of immersion, the T$\rm_c$ slightly decreases to 4.94 K, and the critical current density of TiN films immersed in KOH for 9 days also shows a slight decrease compared to pristine TiN films (see Fig.S3, ESI\dag). Thus, the superconducting properties of TiN are robust in both acidic and alkaline environments. 

\begin{figure*}[t]
	\centering
	\includegraphics[height=15cm]{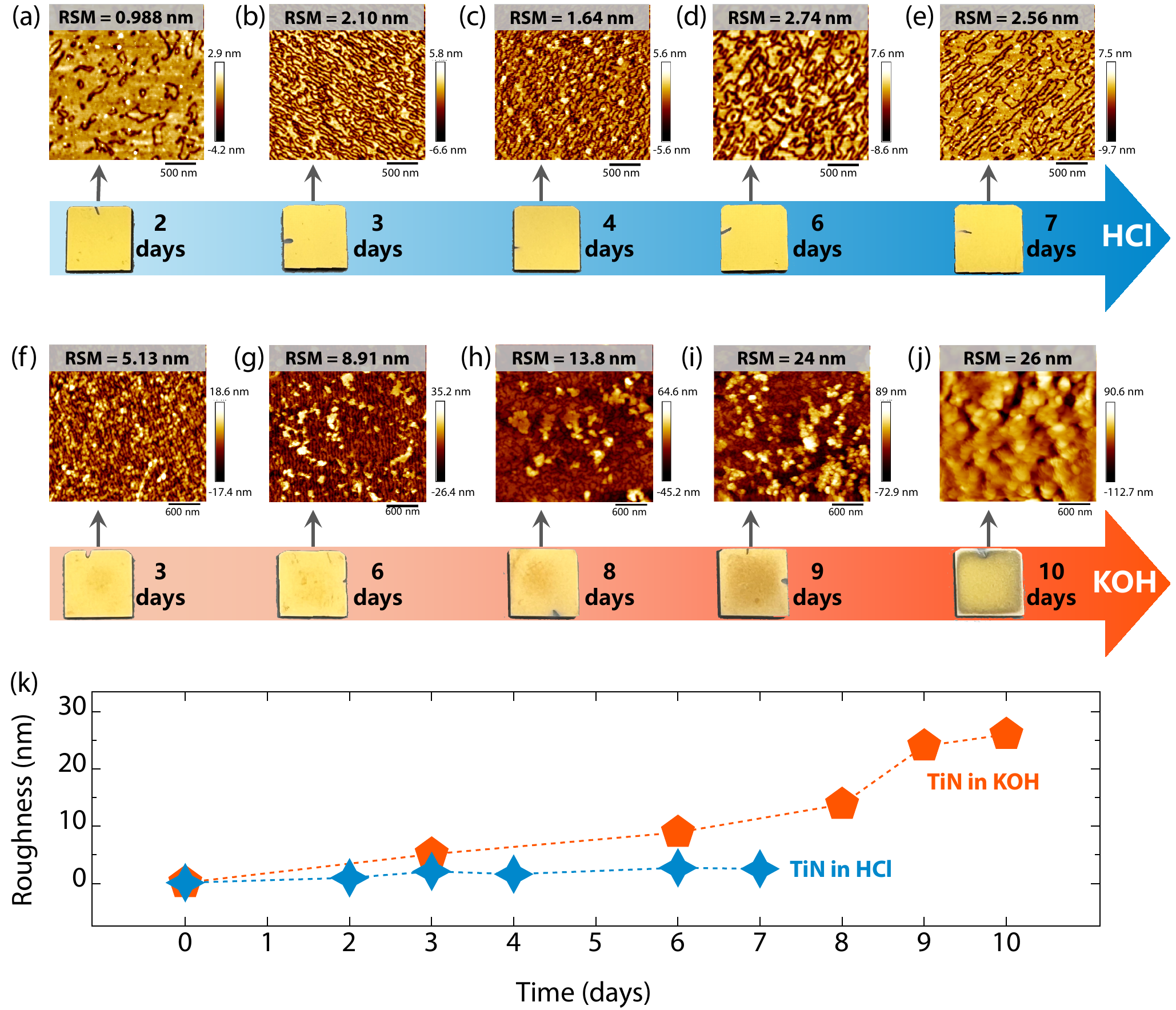}
	\caption{AFM and photographs of TiN films immersed in HCl (a-e, for 2, 3, 4, 6, and 7 days) and KOH (f-j, for 3, 6, 8, 9, and 10 days) solutions. (k) The root mean square roughness of TiN films immersed in HCl and KOH solutions.}
	\label{figure5}
\end{figure*}

\begin{figure*}[t]
	\centering
	\includegraphics[height=10cm]{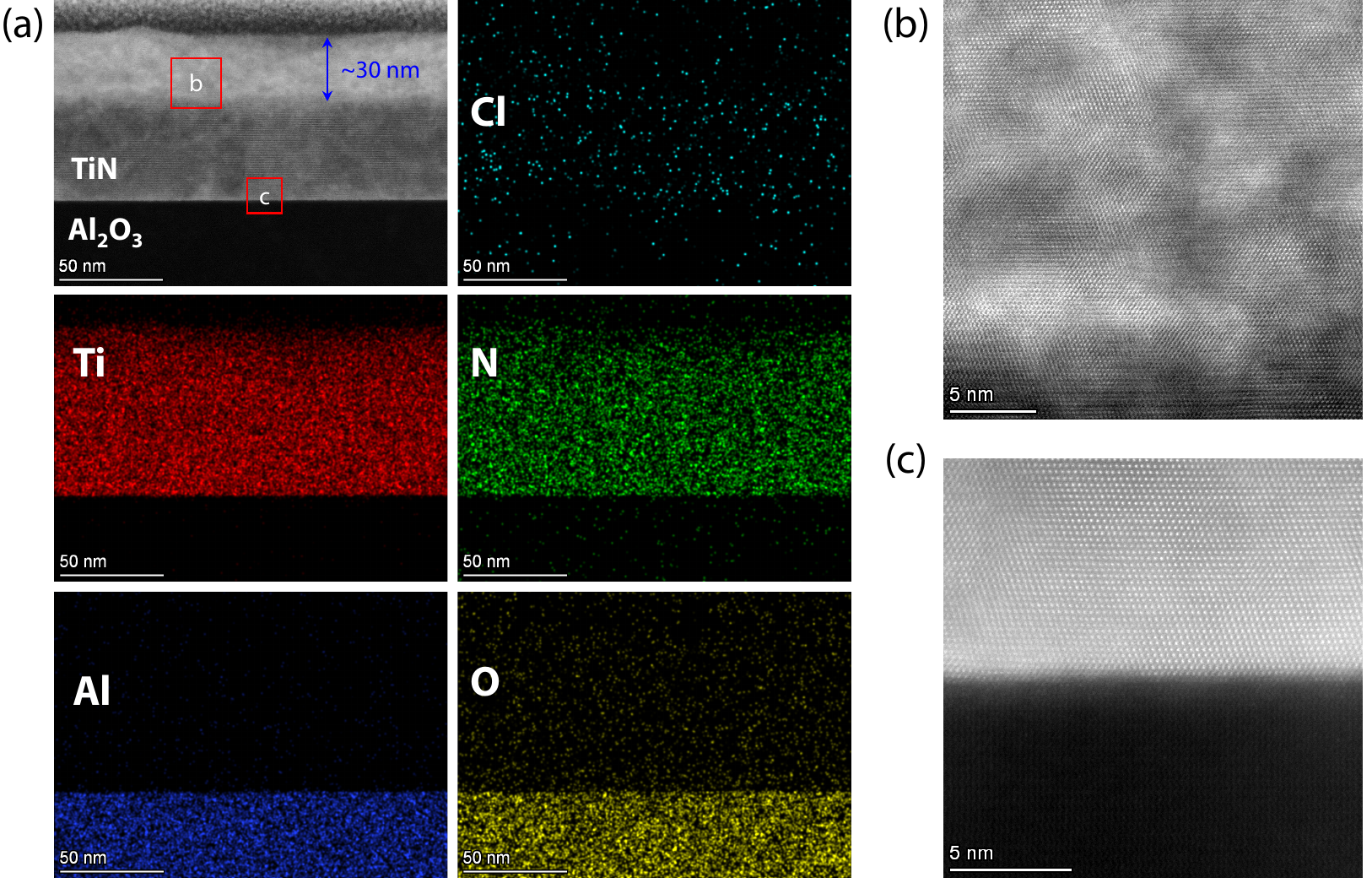}
	\caption{(a) HAADF-STEM image of TiN films on Al$_2$O$_3$ substrates immersed in HCl for 7 days, combined with the EDX mapping of Cl, Ti, N, Al, and O. (b) Enlargement of the brighter area as depicted in (a). (c) Atomic-resolution image of the TiN film at the interface of the film and the substrate as depicted in (a).}
	\label{figure6}
\end{figure*}

In contrast to TiN, the crystal and electronic structures of infinite-layer nickelate and cuprate high-temperature superconductors are sensitive to acidic and alkaline environments. \cite{JAC-2024-Zhao,ACSAMI-2022-Huang,SST-1988-Liu} As shown in Fig.\ref{figure4}, the La$_{0.8}$Sr$_{0.2}$NiO$_2$ films lose their superconductivity rapidly when immersed in acidic solutions, and the same phenomenon presents after 3 hours in an alkaline environment. Additionally, Bi$_2$Sr$_2$CaCu$_2$O$_{8+\delta}$ flakes can be quickly etched in acidic and alkaline solutions,\cite{ACSAMI-2022-Huang} and both HCl and NaOH environments can damage YBa$_2$Cu$_3$O$_{7-\delta}$. \cite{SST-1988-Liu} Meanwhile, despite extensive efforts to grow stable two-dimensional transition metal selenides, the T$\rm_c$ of stable NbSe$_2$ still decreases within 30 minutes  (see Fig.\ref{figure4}). \cite{NM-2019-Lin} Furthermore, the two-dimensional superconducting AuSn$_4$ crystals can corrode in diluted hydrochloric acid. \cite{CM-2020-Shen} Remarkably, the TiN films retain their excellent superconducting properties even after immersion in acidic or alkaline environments for over 7 days.

To understand the surface corrosion processes of superconducting TiN films in acidic and alkaline environments, the evolution of the surface morphology and roughness of TiN films immersed in HCl and KOH for various durations were investigated. Surprisingly, immersion in HCl solution had a negligible effect on the macroscopic morphology of the TiN films, as evident in the photographs in Fig. \ref{figure5}. These films maintained their bright golden color even after 7 days of exposure. The microscopic morphology of TiN films changed with the extension of immersion time in HCl, as shown in Fig.\ref{figure5}(a-e). As seen, the root mean square (RMS) roughness  of TiN films immersed in HCl increased, yet the surface is nano flat, with all values below 3 nm (see Fig.\ref{figure5}(a-e) and (k)). In contrast, the TiN films showed significant corrosion in KOH solution, which becomes worse with extended immersion time (see the photographs in Fig. \ref{figure5}). However, all the TiN films remained remarkably stable on the Al$_2$O$_3$ substrates. Additionally, the corrosion was more pronounced at the center of the samples than at the edges.  As the immersion time in KOH was extended, the surface morphology of TiN films became notably rougher (see \ref{figure5}(f-j) and (k)), with the RSM roughness reaching up to 26 nm after 10 days of immersion (see Fig.\ref{figure5}(j)). This increase in roughness is consistent with the disappearance of thickness fringes in Fig.\ref{figure2}(a) and Fig.\ref{figure3}(a). Therefore, although superconducting TiN films resist both acid and alkali corrosion, the corrosion processes are totally different. Alkaline solutions cause more heavy corrosion of the surface morphology of TiN films, as shown in Fig.\ref{figure5}(k). Meanwhile, the surface morphology of TiN films on Al$_2$O$_3$ substrates was characterized by SEM. As seen in Fig. S5 (ESI\dag), the pristine TiN films show a smooth, homogeneous surface morphology with triangular features, consistent with the features of AFM images of the films (see Fig.\ref{figure1}(e)).  However, after being immersed in HCl and KOH solutions, the worm-like homogeneous morphology emerged (see Fig. S5, ESI\dag). Over time, the surface textures became more pronounced, indicating increased surface roughness.

\begin{figure*}[t]
	\centering
	\includegraphics[height=9cm]{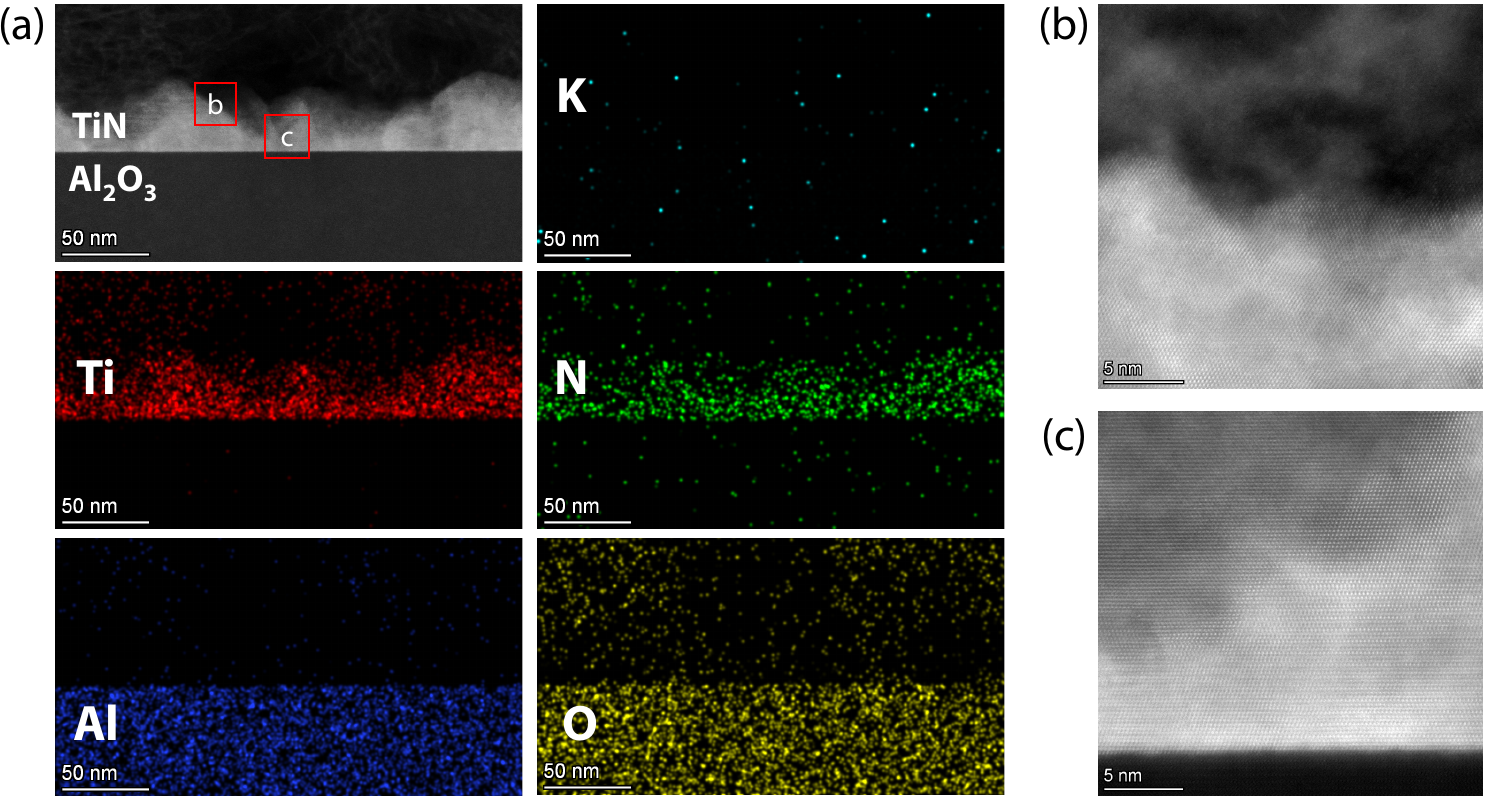}
	\caption{(a) HAADF-STEM image of TiN films on Al$_2$O$_3$ substrates immersed in KOH for 10 days, combined with the EDX mapping of K, Ti, N, Al, and O elements. (b) Enlargement of the top area as depicted in (a). (c) Atomic-resolution image of TiN films at the interface of the film and the substrate as depicted in (a).}
	\label{figure7}
\end{figure*}

To uncover the corrosion processes of superconducting TiN films from the surface to the interior in acidic and alkaline environments, the atomic scale crystal structures of TiN films were characterized by high-resolution STEM.  Figure \ref{figure6} (a) shows the HAADF-STEM images and energy-dispersive X-ray (EDX) mappings of TiN films on Al$_2$O$_3$ substrates immersed in HCl for 7 days. The images reveal minimal chlorine (Cl) infiltration, with the interface of the TiN films and the Al$_2$O$_3$ substrates remaining atomically sharp and free of atomic interdiffusion across the interface. Interestingly, the oxygen content within the film is very low, indicating that acid corrosion did not lead to significant oxidation. Additionally,  the film thickness, which is uniformly within 80 nm, remains unaffected by acid corrosion. A distinct brighter region, approximately 30 nm thick, appears in the TiN films, as shown in Fig.\ref{figure6} (a), possibly due to HCl corrosion. This phenomenon is also observed in TiN films immersed in HCl  for 6 days (Fig.S6, ESI\dag), with the thickness of the brighter region comparable to that of the films after 7 days of immersion (Fig.S6, ESI\dag). The interface between the brighter and darker regions of TiN films is relatively sharp. An enlarged view of the brighter region, shown in Fig. \ref{figure6} (b), reveals numerous defects, yet the films maintain atomic resolution. In contrast, the TiN films near the Al$_2$O$_3$ substrate retain atomic resolution and well-defined characteristics, as seen in Fig. \ref{figure6} (c), demonstrating the high quality of the TiN films. The HAADF-STEM images and EDX mappings of TiN films immersed in HCl for 6 days (Fig.S7, ESI\dag) show results consistent with those of films immersed for 7 days. Therefore, acidic corrosion of TiN films results in significant surface defects without noticeable oxidation, while maintaining high quality within the interior of the films.

For TiN films in KOH solution, there is almost no potassium (K) element detected even after immersion in KOH for 10 days, as shown in the HAADF-STEM images and EDX mappings in Fig. \ref{figure7} (a). Meanwhile, the oxygen concentration in the film area is also very low, indicating no significant oxidation, similar to the films immersed in HCl solution. Interestingly, a notable difference in the corrosion process of TiN films in KOH compared to HCl was the significant decrease in the thickness of the TiN film in KOH.  The brighter mountain-shaped area in Fig. \ref{figure7} (a) is the TiN film, which has a thickness ranging from 20 to 45 nm, consistent with the large roughness shown in Fig. \ref{figure5} (j). Despite the decrease in thickness, the region of the TiN films remains atomic resolution and well-defined, although with some defects (see Fig. \ref{figure7} (b) and (c)). Meanwhile, the interface between the TiN film and the Al$_2$O$_3$ substrate also remains atomically sharp (see Fig. \ref{figure7} (c)). Thus, alkali corrosion can significantly reduce the thickness of the TiN film, leading to pronounced morphological fluctuations, while the TiN film area maintained high crystal quality.

\begin{figure*}[t]
	\centering
	\includegraphics[height=10cm]{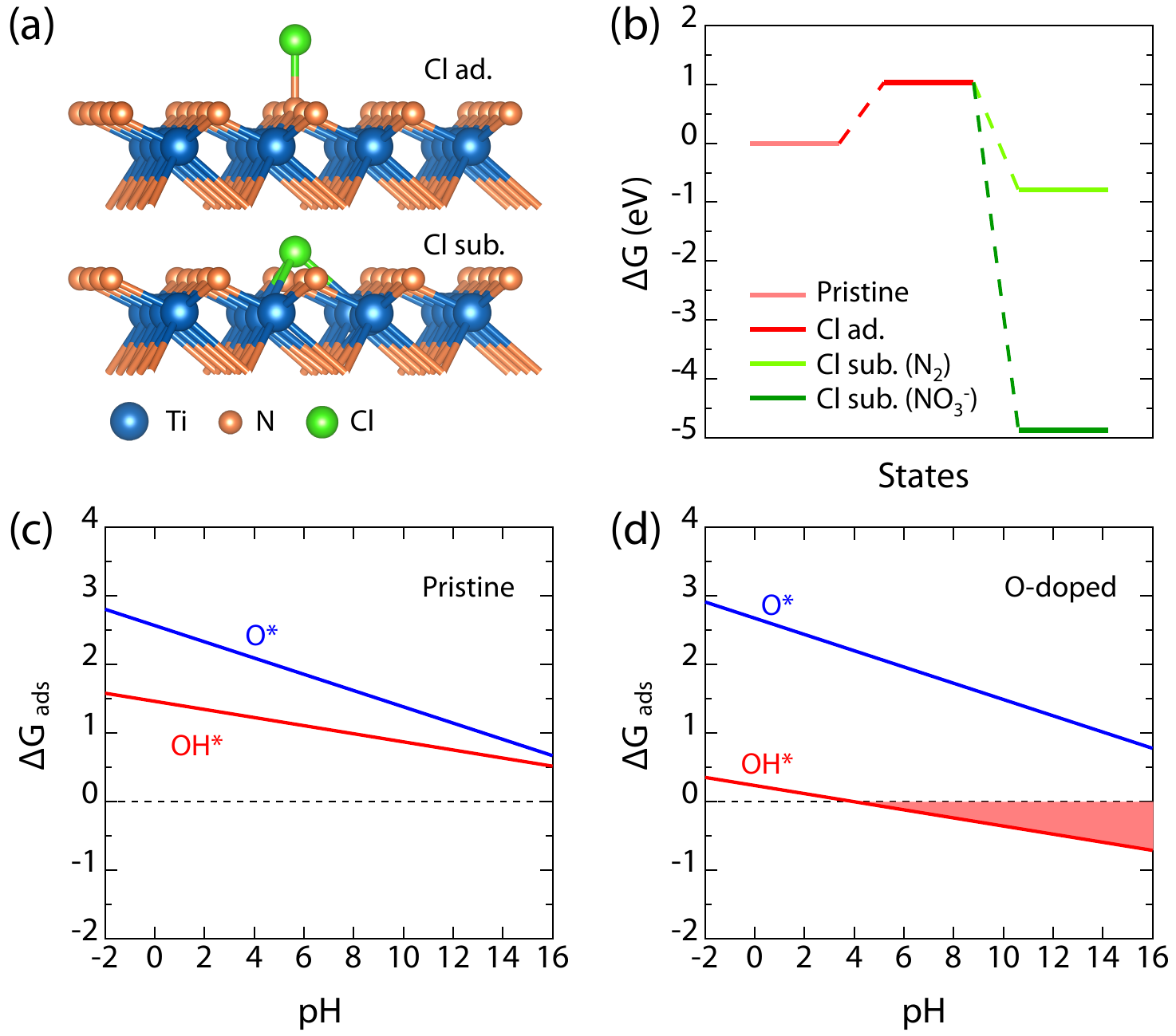}
	\caption{(a) Atomic structural models for TiN (111) surface with Cl adsorbate and substituent. (b) The relative free energies between pristine, adsorbed, and substituted TiN(111) surfaces in acid condition. (c, d) The adsorption free energy of O$^\ast$ and OH$^\ast$ on both pristine and O-doped TiN (111) surfaces.}
	\label{figure8}
\end{figure*}

The above experimental characterizations have shown that the microscopic corrosion behaviors of TiN(111) films are remarkably different in acid and alkaline solutions: (1) In the 1 mol/L HCl solution, the structural integrity of the upper part of TiN(111) film (about 30 nm in Fig. \ref{figure6} (a) and Fig.S7, ESI\dag) is degraded by the gradual inward diffusion of defects created on surface; (2) In the 1 mol/L KOH solution, the upper part of TiN(111) film is gradually eroded away by the oxidative alkaline medium (see Fig. \ref{figure7} (a)), with the left part sustaining the intact lattice structure and superconductivity. To reveal these corrosion mechanisms from an atomistic perspective, we used density-functional theory (DFT) to calculate the electrochemical reactivity of TiN(111) surface in these solutions, for which the detailed DFT parameters and theoretical formula are provided in ESI\dag.

In 1 mol/L HCl solution, according to the widely established corrosion theory,\cite{PMS-2018-Maurice,ARMR-2019-Huang} the promoting effect of corrosive Cl$^{-}$ ion on the creation of structural defects is always the key mechanism. This is consistent with both of the ingression of Cl$^{-}$ into TiN (111) films and the degraded structural integrity after acid corrosion, as observed in our experiment (see Fig. \ref{figure6} and Fig.S7, ESI\dag). The atomic-structure models considered for the adsorption and substitution behaviors of Cl$^{-}$ are shown in Fig. \ref{figure8} (a), and in the latter, the replaced N anion is considered to escape through releasing as the N$_2$ gas or being dissolved into the solution as the NO$^{3-}$ ion. Our DFT results show that the adsorption of Cl$^{-}$ requires an energy barrier of 1.0 eV, however, when Cl$^{-}$ succeeds in substituting a surface N anion, the two reaction paths are both exothermic (see Fig.\ref{figure8} (b)). Thus, the somewhat repulsed adsorption of Cl$^{-}$ ion explains the outstanding corrosion resistance of TiN (111) films as observed in our experiment, while the exothermic defect creation by Cl$^{-}$ ion is responsible for the gradual degraded structural integrity.  In 1 mol/L KOH (pH $\sim$ 14), the adsorption of oxidative substances (i.e., O$^\ast$ and OH$^\ast$  adsorbates) will play the key role in the electrochemical stability of TiN (111) surface. According to the calculated pH-dependent adsorption free energies for O$^\ast$  and OH$^\ast$ (see Fig. \ref{figure8} (c) and \ref{figure8}(d)), it can be seen that OH$^\ast$ is more preferred to O$^\ast$, and these oxidative adsorbates are more stabilized at higher pH values. According to the experimentally observed presence of O impurity in TiN (111) films (see Fig. \ref{figure7}(a)), the effect of O doping on surface adsorption is calculated and found to considerably stabilize the adsorption of OH$^\ast$ , making it exothermic at alkaline conditions (see Fig. \ref{figure8} (d)). It is well known in the corrosion field that it is the exothermic adsorption of OH$^\ast$ that leads to the formation of soluble hydroxides, e.g., Ti(OH)$_4$, Al(OH)$_3$, Cr(OH)$_3$, and Ni(OH)$_2$, \cite{ARMR-2019-Huang,npj-2019-Huang,npj-2023-Huang} on various alloys. Thus, the stabilized OH$^\ast$ here can explain the alkaline erosion of TiN (111) surface observed here (see Fig. \ref{figure7}).

\section{Conclusions}
In conclusion, we have systematically investigated the high resistance of superconducting TiN films against acid and alkali corrosion environments. Surprisingly, the superconducting properties of TiN films remained robust even after immersion in these environments for over 7 days. However, the surface roughness of the TiN films increased with extended immersion time. Furthermore, combining STEM analysis with DFT calculations revealed the corrosion mechanisms: acid corrosion causes numerous defects in the film due to exothermic defect creation by Cl ions substituting the N anion, while alkaline corrosion significantly reduces film thickness resulting from stabilized OH$^\ast$ adsorbates. Despite these surface changes, the interior of the films maintained high crystal quality, leading to their superconducting stability. Our results highlight the unexpected stability and durability of superconducting TiN films, offering potential benefits for their applications in advanced electronics.

\section{Experimental details}

High-quality TiN (111)  films ($\sim$ 80 nm) were synthesized on $\alpha$-Al$_{2}$O$_{3}$ (0001) single crystal substrates (5$\times$5$\times$0.5 mm$ ^{3} $)  by a home-made magnetron sputtering epitaxy system using 2-inch Ti (purity of 99.995\%) target and N$_2$ (purity of 99.999\%) reactive gas. \cite{PRM-2021-Bi,NanoLetter-2024-Bi} The base vacuum pressure of this sputtering system is $\sim$ 3 $\times$10$^{-8}$ Torr. During growth, the N$_2$ pressure was kept at 0.02 Torr with a gas flow of 6.4 sccm, and the substrate temperature was held at 1100 $^\circ$C. The power of the RF generator was 100 W. After growth, the samples were cooled down to room temperature at 25 $^\circ$C per minute in the 0.02 Torr  N$_2$ atmosphere.

The crystal structures of TiN films were characterized by a high-resolution XRD diffractometer (Bruker D8 Discover) with the Cu K$ _{\alpha} $ source ($ \lambda $ = 1.5405 \AA). The surface morphology of TiN films was characterized by atomic force microscopy (Bruker Dimension Icon) and scanning electron microscope (Thermo scientific Verios G4 UC) at room temperature. The atomic-scale HAADF-STEM images and EDX spectroscopy measurements  were collected with an ARM200CF (JEOL, Tokyo, Japan) transmission electron microscope and a Themis Z Double spherical aberration corrected transmission electron microscope. The electrical transport was measured by the Quantum Design Physical Property Measurement System (PPMS) in a van der Pauw geometry. The width of the TiN films is around 0.5 mm for I-V measurements.

\section{DFT Parameters}

The DFT calculations are performed using the Quantum ESPRESSO code package.\cite{JPCM-2009-Giannozzi} The Perdew-Burke-Ernzerhof functional in the generalized gradient approximation is used to describe the electronic exchange and correlation. \cite{PRL-1997-Perdew} The projector augmented wave method \cite{PRB-1994-Blochl} is used to express the electronic wave functions and potentials. The cutoff energies for the wave functions and charge densities are 60 and 600 Ry, respectively. The energetic convergence threshold is 10$^{-6}$ Ry.  The Brillouin zones of the periodic slab models of N-terminated TiN(111) are sampled by a Monkhorst-Pack grid of 3$\times$3$\times$1. \cite{PRB-1976-Monkhorst}  The surfaces of TiN(111) are modeled by using the lattice slab having 7 atomic layers, with the lateral size of 4$\times$4 times of the hexagonal lattice vectors. Thus, the TiN(111) slab has a thickness of $\sim$ 9.6 \AA  ~and a lateral dimension of 11.8$\times$11.8 \AA $^{2}$, with the total atomic number of 112.  The neighboring slab images are separated by a vacuum spacing of $\sim$ 15 \AA ~that can effectively exclude any inter-slab interaction.  To simulate the effect of water solution on the surface properties, an implicit solvation model is considered here, i.e., the self-consistent continuum solvation model. \cite{JCP-2012-Dabo} The van der Waals force \cite{JCP-2010-Grimme} is also considered for the adsorption processes on surface.

\section*{Author Contributions}

Jiachang Bi, Liang-Feng Huang, and Yanwei Cao conceived the project; Zhangyuan Guo, Jiahui Zhang, and Rongjing Zhai prepared and characterized the samples; Min Ge and Qinghua Zhang carried out STEM experiments; You-Qi Zhou, Jin-Tao Ye, and Liang-Feng Huang performed the calculations;  Jiachang Bi wrote the draft; Yanwei Cao and Liang-Feng Huang provided resources and edited the manuscript. All authors discussed the results.

\section*{Conflicts of interest}
There are no conflicts to declare.

\section*{Data availability}
The data supporting this article have been included as part of the Supplementary Information.

\section*{Acknowledgements}
We acknowledge insightful discussions with Tao Chen. This work was supported by the \textquotedblleft Pioneer\textquotedblright \ and \textquotedblleft Leading Goose\textquotedblright \ R\&D Program of Zhejiang (2024C01252(SD2)), the National Key R\&D Program of China (Grant No. 2022YFA1403000), and the Ningbo Science and Technology Bureau (Grant No. 2022Z086). The Supercomputing Center at Ningbo Institute of Materials Technology and Engineering is also acknowledged for providing the computing resources.

\balance

\renewcommand\refname{References}



\newpage

\end{document}